\documentclass[12pt,tightenlines,aps,prd,nofootinbib,superscriptaddress,eqsecnum]{revtex4}

\usepackage{amsmath,amssymb}
\usepackage[mathscr]{euscript}
\usepackage{graphicx}
\usepackage[pdftex]{color}
\usepackage[sort&compress]{natbib}
\usepackage{mathbbol}  

\usepackage[colorlinks=true,linkcolor=blue,filecolor=blue,urlcolor=blue,citecolor=blue,pdftex=true,plainpages=false]{hyperref}

\usepackage{algorithm}
\usepackage{algpseudocode}

\usepackage{xcolor}
\newcommand{\ReTr}{\mathbb{Re Tr}}
\newcommand{\Tr}{\mathbb{Tr}}

\begin{document}

\title{Diffusion model for SU($N$) gauge theories}

\author{Javad Komijani}
\email{jkomijani@ethz.ch}
\author{Marina K. Marinkovic}
\author{Lara Turgut}

\affiliation{Institute for Theoretical Physics, ETH Zurich, 8093 Zurich, Switzerland}

\date{\today}

\begin{abstract}
Implicit score matching provides a computationally efficient approach for training
diffusion models and generating high-quality samples from complex distributions.
In this work, we develop a score-matching framework for SU($N$) lattice gauge
theories, which can be extended to other Lie groups. We apply the method to
SU(3) gauge configurations with the Wilson gauge action in two and four dimensions
and assess the quality of the generated samples by comparison with Hybrid Monte
Carlo (HMC) simulations. We show that the diffusion models can be successfully
trained and applied for sampling the Wilson gauge action.
For large values of inverse coupling, accurate reverse-time integration requires
predictor--corrector schemes, for which we introduce a corrector based on
Hamiltonian molecular dynamics.
While the corrector significantly improves sampling quality, it also increases the computational cost.
We outline several strategies for improving sampling efficiency.
\end{abstract}

\maketitle

\section{Introduction}
\label{sec:intro}

Lattice gauge theory enables precise calculations of a wide range of observables in quantum chromodynamics (QCD), relying on the efficient sampling of gauge-field configurations using methods such as Hybrid Monte Carlo (HMC)~\cite{Duane:1987de}. However, HMC suffers from critical slowing down as the lattice spacing is reduced, leading to long autocorrelation times that in practice limit the achievable precision of numerical results for fixed computational resources.

Recent advances in generative modeling have introduced alternative methods to tackle the sampling problem in lattice gauge theory, with diffusion models (DM)~\cite{Zhu:2025pmw, Aarts:2026zzr}
and normalizing flows~\cite{Kanwar:2020xzo, Boyda:2020hsi, Kanwar:2024ujc, Bacchio:2022vje} as two promising approaches.
In this paper, we apply diffusion models to lattice gauge theory with the SU(3) gauge group in two and four spacetime dimensions.

Diffusion models~\cite{Sohl-Dickstein:2015, Song:2019Langevin, Ho:2020denoising} are built on the idea that one can define a gradual noise-injection process that transforms the data distribution toward a simple reference distribution at a final diffusion time (typically a Gaussian distribution, or a uniform distribution for compact groups).
The crucial point is that this process can be reversed: starting from the reference distribution, one reconstructs samples for the original data distribution by integrating a reverse-time dynamics.
The reverse process is fully determined by the time-dependent \emph{score function}, i.e., the gradient of the logarithm of the probability density of the diffused distribution at each time.
The main task in diffusion modeling is to learn this score function from data.

A remarkable feature of diffusion models is that the learned score function can be exploited through several closely related sampling formulations.
for a given forward diffusion, there exists a family of reverse-time dynamics that all have the same marginal distributions at each diffusion time~\cite{Vega:2025hgz}.
The reverse-time dynamics can be expressed as a stochastic differential equation (SDE), which can be simulated using standard discretizations such as Euler--Maruyama, or equivalently as a deterministic ordinary differential equation (ODE), which can be solved using conventional ODE integrators.
Moreover, these approaches can be combined in predictor--corrector schemes, where a predictor step advances the configuration backward in time (using either an SDE or ODE update), while a corrector step applies a Markov transition that improves
sampling at fixed diffusion time.
This flexibility is particularly appealing for lattice gauge theories, where the stability and efficiency of reverse-time integration can depend on the coupling, and it motivates the integration strategies explored in this work.

In the context of lattice field theory, diffusion models have recently been explored in a range of two-dimensional models, including $\phi^4$ scalar field theory~\cite{Wang:2023exq}, U(1) gauge theory~\cite{Zhu:2025pmw}, and non-Abelian gauge theory~\cite{Aarts:2026zzr}, as well as for theories with complex actions~\cite{Aarts:2025lpi}.
In addition, ref.~\cite{Vega:2025hgz} applies diffusion models to $\phi^4$ scalar and U(1) gauge theories in two dimensions using a modified implicit score-matching objective, augmented by incorporating the exact score of the target theory to constrain the learned score at the initial diffusion time.

For a concise review of diffusion models and their applications to lattice scalar field theory and abelian gauge theory, we refer the reader to Ref.~\cite{Vega:2025hgz}.
In this work, we extend diffusion-model sampling to non-Abelian lattice gauge theories
and develop an implicit score-matching framework for SU($N$) gauge groups.
We also propose a predictor–corrector sampler with a corrector designed based on Hamiltonian molecular dynamics (HMD)~\cite{Gottlieb:1987mq}; see ref.~\cite{Neal:2011mrf} for a review of Monte Carlo simulation with Hamiltonian dynamics.
Furthermore, we introduce gauge-equivariant networks, including a U-Net–style architecture~\cite{Ronneberger:2015unet} suitable for large lattices.
The proposed algorithm is applied to the SU(3) gauge group with the Wilson action in two and four dimensions, and it can easily be generalized to other Lie groups.

The paper is organized as follows.
In section~\ref{sec:DM-simulation}, we formulate diffusion models on Lie groups for lattice gauge fields, detailing the forward and reverse-time dynamics, present our implicit score-matching training objective.
We then illustrate the proposed method on the SU(2) and SU(3) toy models. In section~\ref{sec:gauge-theory}, we apply this approach to SU(3) lattice gauge theory: we introduce gauge-equivariant architectures for constructing score functions, describe Langevin- and HMD-based sampling methods, present simulations in two and four dimensions, and compare the resulting generative procedure against HMC.
We conclude in section~\ref{sec:conclusion} with a summary, discussion, and outlook.
\section{Diffusion Models on Group Manifold}
\label{sec:DM-simulation}

In this section, we formulate diffusion models for non-Abelian lattice gauge theories.
We define a forward diffusion process that gradually transforms an initial structured distribution into the uniform distribution defined by the Haar measure.
We then present the corresponding Fokker--Planck (FP) equation, which governs the evolution of the probability density along diffusion time and depends on a time-dependent score function that encodes the drift necessary to reverse the diffusion. 
Building on this, we introduce the family of reverse-time dynamics that form the basis of different sampling methods.
To train the model efficiently, we present an implicit score-matching scheme, which bypasses the need to evaluate the generally intractable likelihood of the diffused distribution.
Finally, we illustrate the formalism through concrete examples using SU(2) and SU(3) random matrix models, while the application to full SU(3) lattice gauge theory is deferred to the next section.

\subsection{Forward process}

We denote the diffusion time by $t\in[0,1]$, the initial Lie-group element by $U_0$, and the diffused element at time $t>0$ by $U_t$.
In the lattice gauge theory setting, $U_t$ corresponds to a link variable and thus depends on the lattice site $x$ and direction $\mu$, i.e.\ $U_t \equiv U_t(x,\mu)$.
To simplify the notation, we suppress the explicit $(x,\mu)$ dependence whenever it is not essential.

We define the forward diffusion process as a left-multiplicative random walk on the group manifold, $U_t = K_{t, 0}\, U_0$,
where $K_{t, 0}$ is defined \emph{formally} using the time-ordering operator $\mathcal{T}$ as
\begin{equation}
   K_{t, t'} = \mathcal{T}\exp\!\left(\int_{t'}^{t} \sigma(\tau) dW_\tau\right),
\quad t \ge t'\,,
\end{equation}
with $\sigma(\tau)$ a scalar noise schedule.
Here $dW_\tau$ is a Lie-algebra–valued increment expanded as
\begin{equation}
    dW_\tau = dW^a_\tau\, T^a\,,
\end{equation}
where the coefficients $dW^a_\tau$ are independent Wiener increments, and $\{T^a\}$
are generators of the corresponding Lie algebra.
The evolution remains on the group, and it satisfies
\begin{equation}
    K_{t, t''} = K_{t, t'} K_{t', t''}\,, \qquad t\ge t' \ge t''\,,
\end{equation}
which implies $U_t = K_{t, t'} U_{t'}$.

In practice, we divide the diffusion-time interval into $M$ steps and approximate $K_{t, t'}$ by a product of $M$ group-valued random matrices:
\begin{align}
  \left\{
  \begin{aligned}
    U_t &= K_{t, t'}\, U_{t'}\,,  \\
    K_{t, t'} &= \prod_{m=0}^{M-1} \exp\left(\sigma_\text{cum}(t_m,\, t_{m+1})\, \eta(t_m)\right),
  \end{aligned}
  \right.
  \label{eq:diffusion:U_t}
\end{align}
where $t_m = t' + mh$ with $h = (t-t')/M$.
The product yields the formal time-ordered exponential as $M\to\infty$.
Here,
\begin{equation}
    \eta(t) = \eta^a(t)\, T^a\,,
\end{equation}
is a Gaussian random element of the Lie algebra, whose independent components $\eta^a(t)$ are drawn from a normal distribution with zero mean and unit variance.
The scalar function $\sigma_{\mathrm{cum}}(a,b)$ controls the accumulated noise between diffusion times $a$ and $b$, defined by
\begin{equation}
    \sigma^2_\text{cum}(a, b) = \int_a^b dt\, \sigma^2(t).
\end{equation}
Throughout this work we employ the noise schedule
\begin{equation}
    \sigma(t) = \frac{\sigma_0}{\sqrt{1 - t + \varepsilon}}\,,
    \label{eq:noise:schedule}
\end{equation}
where $\varepsilon$ is a small regulator that prevents the variance from diverging at $t=1$.
Unless otherwise specified, we use $\sigma_0 = 1 / \log(10)$ and $\varepsilon=10^{-8}$.

The diffusion process defined in Eq.~\eqref{eq:diffusion:U_t}, together with the noise schedule in Eq.~\eqref{eq:noise:schedule}, drives the distribution of $U_t$ toward the uniform distribution on the Lie group as $t\to 1$ and, in the lattice gauge-theory setting, decorrelates neighboring links.
Therefore, the diffusion process provides an stochastic interpolation between structured data at $t=0$ and a reference distribution at $t=1$.
The form of this interpolation is influenced by the choice of the noise schedule.

As a numerical illustration, we apply the forward diffusion to toy datasets consisting of SU(2) and SU(3) random matrices. The matrices are drawn from the probability density function (PDF)
\begin{equation}
    \rho_0(U) \propto \exp\!\left(\frac{\beta}{N} \ReTr(U)\right),
\end{equation}
with $\beta=6$ and $N=2, 3$ for SU(2) and SU(3), respectively.
Throughout this paper, probabilities on the group are defined with respect to the Haar measure.
The datasets are generated using a normalizing-flow approach implemented in the \texttt{normflow} package~\cite{Komijani:2025normflow, Komijani:2025yjz, Boyda:2020hsi}.

\begin{figure}
    \centering
    \includegraphics[width=0.48\linewidth, trim=2cm 0.2cm 2cm 1.5cm, clip]{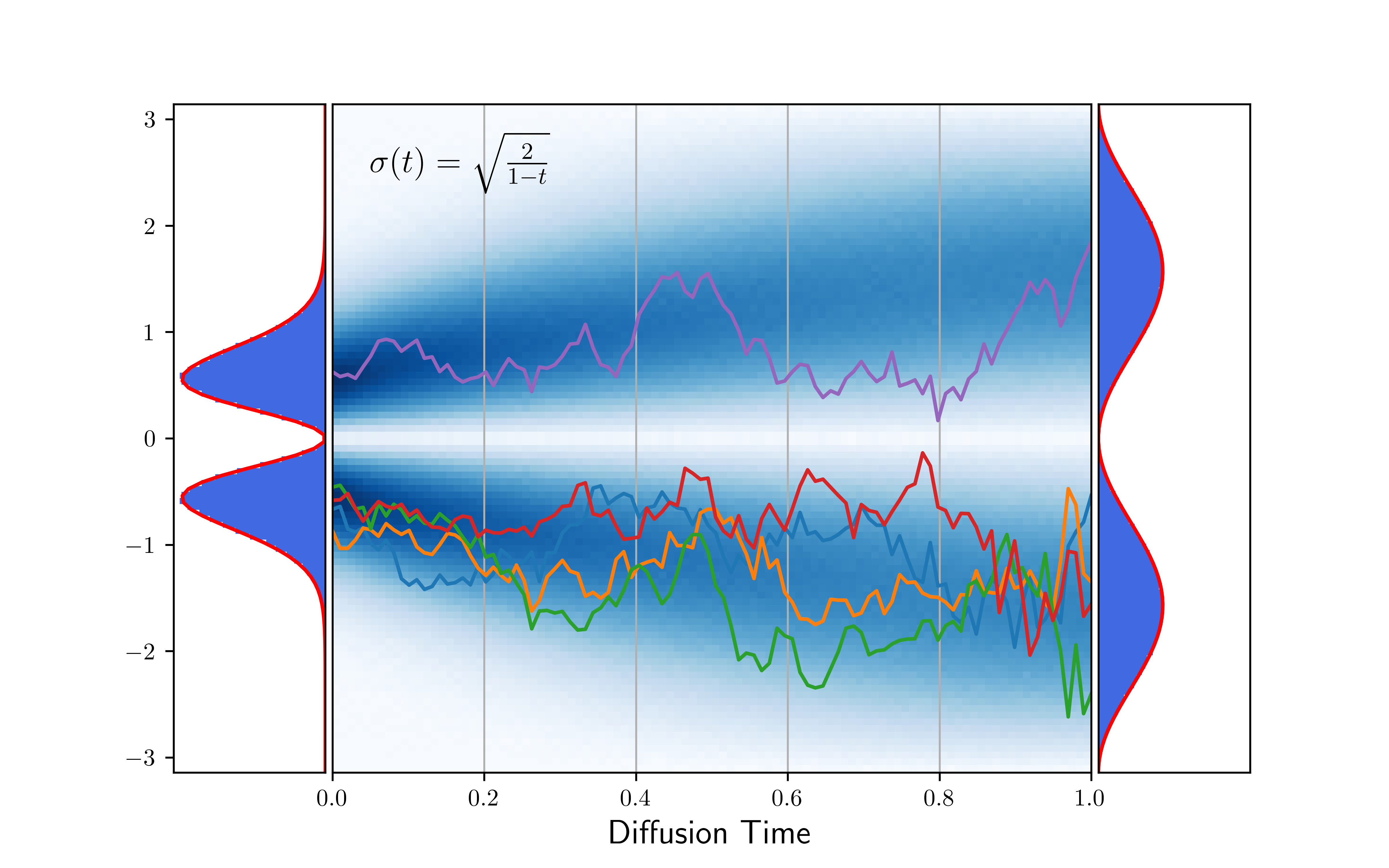}
    \includegraphics[width=0.48\linewidth, trim=2cm 0.2cm 2cm 1.5cm, clip]{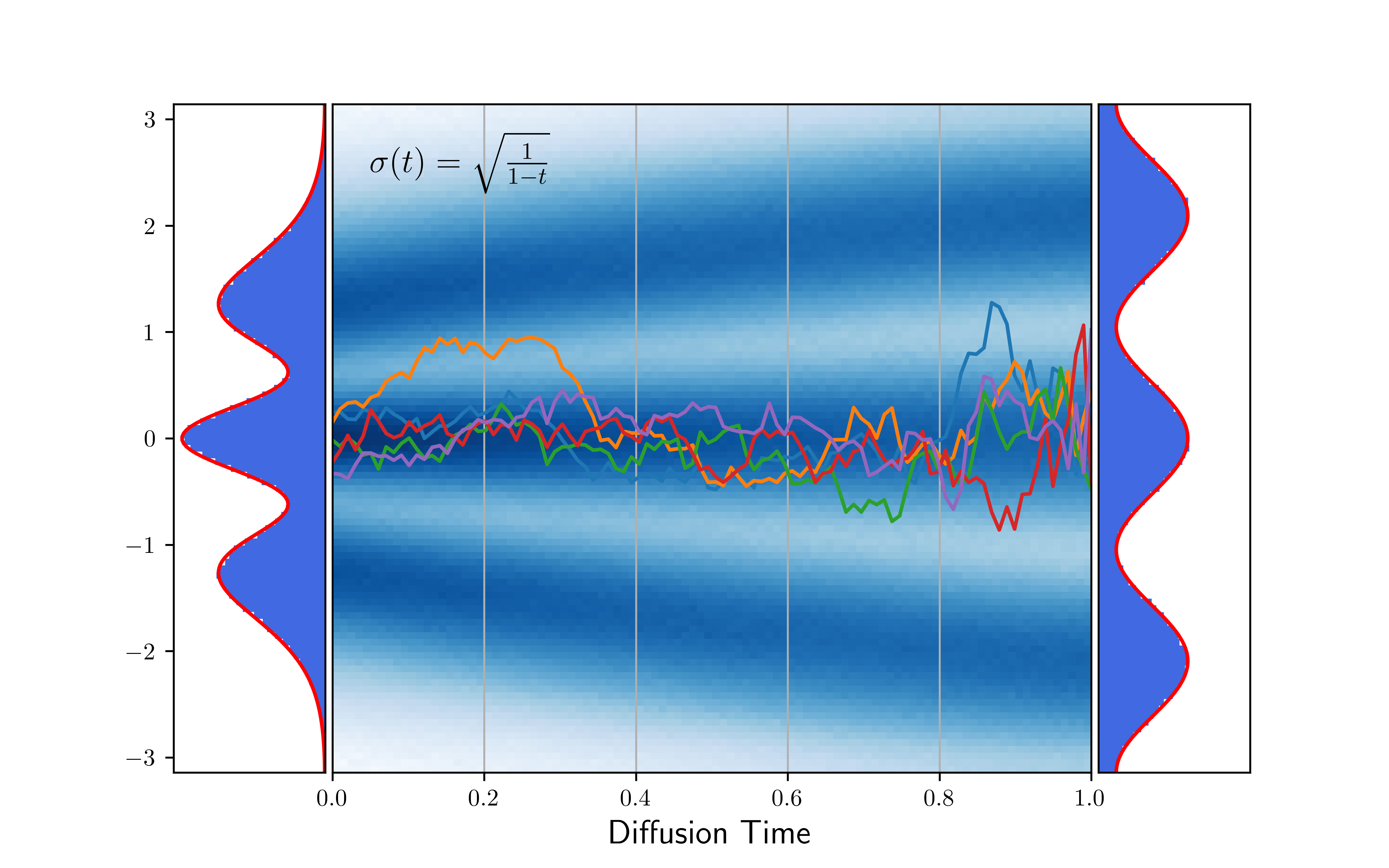}
    \caption{The evolution of the eigenangles of SU($N$) matrices as a function of diffusion time according to the diffusion process
    defined in eq.~\eqref{eq:diffusion:U_t}.
    Left: SU(2) matrix model. Right: SU(3) matrix model.
    The lines represent five sample trajectories of the eigenangles stochastically evolving towards $t = 1$.
    }
    \label{fig:SU_matrix_eigval_diffusion}
\end{figure}

In figure~\ref{fig:SU_matrix_eigval_diffusion}, we show examples of diffusion trajectories generated according to eq.~\eqref{eq:diffusion:U_t} for both datasets.
Since the probability density depends only on $\ReTr(U)$, it is fully determined by the eigenvalues of the group element $U$.
For SU$(N)$, these eigenvalues lie on the unit circle and can be parametrized by their angles, which we refer to as eigenangles.
Figure~\ref{fig:SU_matrix_eigval_diffusion} shows the evolution of the histogram of eigenangles as a function of diffusion time, together with the trajectories of five randomly selected samples.
As expected, the distribution smoothly interpolates between the initial structured distribution at $t=0$ and the uniform (Haar) distribution at $t=1$.

The probability density function of $U_t$ evolves from the initial distribution, $\rho_0$, toward the uniform distribution at $t=1$, which corresponds to vanishing $\beta$. 
We denote by $\rho_t(U_t)$ the probability density of the diffused samples at diffusion time $t$. 
While $\rho_t$ generally depends on the noise schedule $\sigma(t)$, we suppress this dependence in the notation for simplicity.

The time evolution of $\rho_t$ is governed by the corresponding Fokker--Planck (FP) equation,
\begin{equation}
    \partial_t \rho_t(U)
    = \frac{\sigma^2(t)}{2}\, \partial^a \partial^a \rho_t(U),
    \label{eq:FP:Lie:no-drift}
\end{equation}
where $\partial^a$ denotes the left-invariant derivatives on the group manifold.
This equation describes diffusion on the Lie group and contains no drift term, reflecting the fact that the forward process consists purely of stochastic noise. 
A detailed derivation of eq.~\eqref{eq:FP:Lie:no-drift}, along with a full explanation of the notation, is provided in Appendix~\ref{apx:Diffusion-process}.
Equation~\eqref{eq:FP:Lie:no-drift} can be viewed as a special case of the more general FP equation~\eqref{eq:FP:Lie}, obtained by setting the drift term to zero.

Before proceeding, we make a brief remark regarding the choice of the time step $h$ in eq.~\eqref{eq:diffusion:U_t}. 
For the diffusion trajectories shown in figure~\ref{fig:SU_matrix_eigval_diffusion}, we use $h = 0.01$. 
With this choice, $U_t$ evolves smoothly toward the uniform distribution with respect to the Haar measure as $t\to 1$, reproducing the expected behavior of the forward process. 
In practice, the step size $h$ can be adjusted depending on the desired numerical accuracy and computational efficiency. Under suitable conditions, larger values of $h$ may be used without compromising the convergence of the kernel toward the uniform distribution with respect to 
the Haar measure. Further discussion of this flexibility is provided below in section~\ref{sec:score-matching}.

\subsection{Reverse process}

The diffusion process defined in eq.~\eqref{eq:diffusion:U_t} is reversible in the sense that evolving samples of $U_{t_1}$ forward from time $t_1$ to $t_2 > t_1$, and then evolving them backward from $t_2$ to $t_1$, leaves the probability distribution of $U_{t_1}$ unchanged.
In Appendix~\ref{apx:Diffusion-process}, we derive the corresponding reverse-time process:
\begin{align}
  \left\{\begin{aligned} U_t &= K_{t, t'} U_{t'}\,, \\
    K_{t, t'} &= \prod_{m=0}^{M-1}
    \exp\left(-\frac{1}{2}\left(dt\, \sigma^2_{t_m} - |dt| \tilde{\sigma}^2_{t_m}\right)
    T^a \partial^a \log \rho_{t_m}(U_{t_m}) + \sqrt{|dt|} \tilde\sigma_{t_m} \eta(t_m)\right), \quad t < t'\,,
  \end{aligned} \right. \label{eq:denoising:U_t}
\end{align}
where $t_m = t' + m dt$ with $dt=(t - t')/M$ and $M$ taken large.
Remarkably, in the special case $\tilde \sigma = 0$, the stochastic term vanishes and the reverse process reduces to a deterministic flow.
In the continuous-time limit $dt\to 0$, this can be interpreted as a continuous normalizing flow:
\begin{equation}
   \frac{dU_t}{dt} = - \frac{1}{2} \sigma^2(t)\, T^a \partial^a \log \rho_t(U_t)\, U_t\,.   \label{eq:denoising:U_t:ODE}
\end{equation}

Both the stochastic and deterministic reverse-time dynamics are governed by the time-dependent score $T^a \partial^a \log \rho_t(U_t)$, i.e. the gradient of the log-density of the diffused distribution at time $t$.
Because this quantity is generally intractable, we approximate it with a learned score model, as described in the section~\ref{sec:score-matching}.

\subsection{Score matching}
\label{sec:score-matching}

We learn the score by introducing a parametric model $\mathbf{s}_\theta(t, U_t)$ and training it to approximate the true score $T^a \partial^a \log \rho_t(U_t)$ of the diffused data at time $t$.
This is achieved by minimizing a suitable score-matching objective, e.g., a time-weighted expectation of the squared difference between the estimated and true scores.
By accurately estimating the true score, the model learns the information necessary to reverse the diffusion process and generate samples starting from uniform noise at $t=1$.

Following ref.~\cite{hyvarinen:2005implicit}, the score-matching objective can be rewritten in terms of conditional distributions as
\begin{equation}
    \int dx\, p(x)
    \left\lVert \mathbf{s}_\theta(x) - \nabla \log p(x)\right\rVert^2 
    =
    \int dx dx_0 \, p(x, x_0)
    \left\lVert \mathbf{s}_\theta(x) - \nabla \log p(x|x_0)\right\rVert^2
    + \text{const}.
\end{equation}
The constant appearing on the right-hand side is independent of the model parameters $\theta$ and can be ignored during optimization.
This framework of aligning the model score with the gradient of the conditional log-probability is referred to as \emph{implicit}
score matching~\cite{Vincent:2011score}.

For diffusion processes driven by Wiener noise on $\mathbb{R}^n$, the conditional distribution $p(x|x_0)$ is known analytically, which makes implicit score matching particularly convenient.
In this setting, the conditional score $\nabla \log p(x|x_0)$ can be computed in closed form, allowing the model $s_\theta(x)$ to be trained directly against it.
We now argue that the same principle can be extended to diffusion processes defined on Lie groups such as SU($N$).

To simplify score matching on the group, we introduce an auxiliary process $\Gamma_t$ defined by the infinitesimal update
\begin{equation}
    d\Gamma_t = \log(U_{t+dt} U_t^\dagger)\,,
    \label{eq:dGamma}
\end{equation}
where the logarithm maps group elements to the Lie algebra.
By restricting to the principal branch of the logarithm and applying to the forward diffusion in Eq.~\eqref{eq:diffusion:U_t}, the evolution of $\Gamma_t$ becomes
\begin{equation}
   d\Gamma_t \Big|_\text{forward} = \sigma(t) dW_t\,.
   \label{eq:SDE:Gamma_t:forward}
\end{equation}
Hence, $\Gamma_t - \Gamma_0$ follows a Gaussian process in the Lie algebra, whose components in a generator expansion have variance $\sigma_\text{cum}^2(0,t)$.
Setting $\Gamma_0 = \log U_0$,
the process $\Gamma_t$ encodes the evolution of the group-valued state $U_t$
in the algebra space.
Note that the state $U_t$ is recovered from $\Gamma_t$ through the time-ordered exponential
\begin{equation}
    U_t =
    \mathcal{T}\exp\!\left(\int_0^t d\Gamma_s\right) U_0,
\end{equation}
rather than a simple exponential of $\Gamma_t$.
The evolution can be reversed as
\begin{equation}
    U_0 =
    \mathcal{T}^{-1}\exp\!\left(\int_t^0 d\Gamma_s\right) U_t,
\end{equation}
where $\mathcal{T}^{-1}$ denotes inverse time ordering.

This construction allows score matching to be performed in the Lie algebra, where the conditional score is available in closed form, making implicit score matching directly applicable to Lie-group diffusion processes.

The reversed SDE of $\Gamma_t$ can be derived in two ways:
by reversing \eqref{eq:SDE:Gamma_t:forward}, or by applying \eqref{eq:dGamma} to the reversed process in \eqref{eq:denoising:U_t}.
These approaches yield
\begin{align}
    \left\{
    \begin{aligned}
    d\Gamma_t \Big|_\text{reverse,1}
    &=- dt \frac{\sigma^2 + \tilde\sigma^2}{2} \nabla \log q_t(\Gamma_t)
    + \tilde\sigma dW_t\,, \\
    d\Gamma_t \Big|_\text{reverse,2}
    &= - dt \frac{\sigma^2 + \tilde\sigma^2}{2} T^a \partial^a \log \rho_t(U_t)
    + \tilde \sigma dW_t\,,
    \end{aligned}
    \right.
    \label{eq:diffusion:SDE:rev:Gamma_t}
\end{align}
where $q_t(\Gamma_t)$ denotes the probability density of $\Gamma_t$, and $\rho_t(U_t)$ is the density of corresponding $U_t$.
The drift terms in these two equations are not, in general, identical.
This reflects the fact that, for non-Abelian groups and finite $t$, there is no unique mapping from $\Gamma_t$ to $U_t$.
The two reverse dynamics should indeed be understood in a stochastic sense:
the first preserves the distribution of $\Gamma_t$ under forward–reverse composition, while the second preserves the distribution of $U_t$ when updated via the exponential of $d\Gamma_t$.

We do not provide a rigorous proof, but argue heuristically that the two SDEs
in eq.~\eqref{eq:diffusion:SDE:rev:Gamma_t} are stochastically equivalent, in the sense that both can be used to describe the reverse-time evolution of $U_t$.
This follows from the fact that both are determined by the relation between the group increment $U_t U_{t+dt}^\dagger$ and the Lie-algebra increment $d\Gamma_t$, and therefore encode the same infinitesimal dynamics of $U_t$ in reverse time.

To illustrate this connection, consider the Abelian case.
Here, time ordering is trivial and $\Gamma_t = \log U_t$.
The conditional distribution $\rho_t(U_t \mid U_0)$ is a multivariate wrapped Gaussian, whose gradient can be written as
\begin{equation}
    T^a \partial^a \rho_t(U_t \mid U_0) \propto
    - \sum_{\mathbf{n}} \frac{\Gamma_t - \Gamma_0 + A_{\mathbf{n}}}{\sigma^2_{\mathrm{cum}}(0,t)}
    \exp\!\left(
    -\frac{\mathbb{Tr}\, (\Gamma_t - \Gamma_0 + A_{\mathbf{n}})(\Gamma_t - \Gamma_0 + A_{\mathbf{n}})^\dagger}
    {2 \sigma^2_{\mathrm{cum}}(0,t)}
    \right),
    \label{eq:periodic_gaussian}
\end{equation}
where the sum runs over all branches of the logarithm, and $A_{\mathbf{n}}$ accounts for the corresponding shifts.
An integral over $U_t$ is equivalent to an integral over $\Gamma_t$ restricted to a single branch of the logarithmic map.
Equivalently, this can be expressed as an integral over $\Gamma_t$ on the full algebra with a non-wrapped Gaussian.
Using this observation, we can rewrite the integral appearing in implicit score matching as
\begin{align}
  \int dU_t\, \mathbf{s}_\theta(t, U_t)\, T^a \partial^a \rho_t(U_t \mid U_0)
  &=
  \int d\Gamma_t\, \mathbf{s}_\theta(t, U_t)\,
  \frac{\Gamma_t - \Gamma_0}{\sigma^2_{\mathrm{cum}}(0,t)}
  \exp\!\left(
  -\frac{\mathbb{Tr}\, (\Gamma_t - \Gamma_0)(\Gamma_t - \Gamma_0)^\dagger}
  {2 \sigma^2_{\mathrm{cum}}(0,t)}
  \right)
  \nonumber\\
  &=
  \int d\Gamma_t\, \mathbf{s}_\theta(t, U_t)\, \nabla q_t(\Gamma_t \mid \Gamma_0)\,.
  \label{eq:Abelian:score-equivalence}
\end{align}
This identity shows that, in the Abelian case, score matching on the group reduces exactly to score matching in the Lie algebra.

We now turn to the general non-Abelian case, where the arguments leading to eq.~\eqref{eq:Abelian:score-equivalence} are not valid.
Nevertheless, based on our heuristic connection between the two SDEs
in eq.~\eqref{eq:diffusion:SDE:rev:Gamma_t}, we train the model $\mathbf{s}_\theta(t, U_t)$ to match the algebra-space score rather than the group-space score directly.
Within the implicit score-matching framework, this reduces to matching the model to the conditional score $\nabla \log q_t(\Gamma_t \mid \Gamma_0)$.

Since the forward diffusion in $\Gamma_t$ is Euclidean, the conditional distribution is
Gaussian with score
\begin{equation}
    \nabla_{\Gamma_t} \log q_t(\Gamma_t \mid \Gamma_0)
    =
    - \frac{\Gamma_t-\Gamma_0}{\sigma_\text{cum}^2(0,t)}.
\end{equation}
We can then define the score-matching objective
\begin{equation}
    \int dt\, dU_0\, \rho_0(U_0)\, \mathbb{E}\left[\lambda(t)
    \left\lVert \mathbf{s}_\theta(t, U_t) + \frac{\Gamma_t - \Gamma_0}{\sigma^2_\text{cum}(0, t)}\right\rVert^2\right],
    \label{eq:loss}
\end{equation}
where $\lambda(t)$ is a time-dependent weight, which we set
to $\lambda(t) = \sigma_\text{cum}(0, t)$.

To minimize \eqref{eq:loss}, we proceed as follows.
We begin with a batch of samples at the initial time, denoted by $U_0$.
For each sample, we draw a time $t$ uniformly from the interval $[0, 1]$
and compute $U_t$ using the forward diffusion process \eqref{eq:diffusion:U_t}
with a fixed $M$.
Both $t$ and $U_t$ are then fed into the model score function.
Then, using the forward diffusion \eqref{eq:diffusion:U_t} together with the transformation \eqref{eq:dGamma}, we obtain the expression
\begin{equation}
   \Gamma_t = \Gamma_0 + \sum_{m=0}^{M-1} \sigma_\text{cum}(t_m,\, t_{m+1})\, \eta(t_m)\,,
\end{equation}
which is used in the score-matching objective defined in eq.~\eqref{eq:loss}.
Finally, we average the expression inside the square brackets over the batch, thereby providing a procedural definition of the expectation value in eq.~\eqref{eq:loss}.

Computing $U_t$ requires generating and multiplying $M$ random group elements, and in
principle one expects $M$ to be large in order to accurately approximate the underlying
continuous-time evolution. This can become a computational bottleneck in our score-matching scheme. However, our
empirical investigations indicate that, in the setting considered in this work, the
relevant statistical properties of the process are already captured with surprisingly
small values of $M$. In particular, we find that $M=4$ is sufficient to approximate both $U_t$ in all experiments reported here.

\subsection{Toy models: application to SU(2) and SU(3) matrix models}

\begin{figure}
    \centering
    \includegraphics[width=0.48\linewidth, trim=0.5cm 0.2cm 1cm 0.5cm, clip]{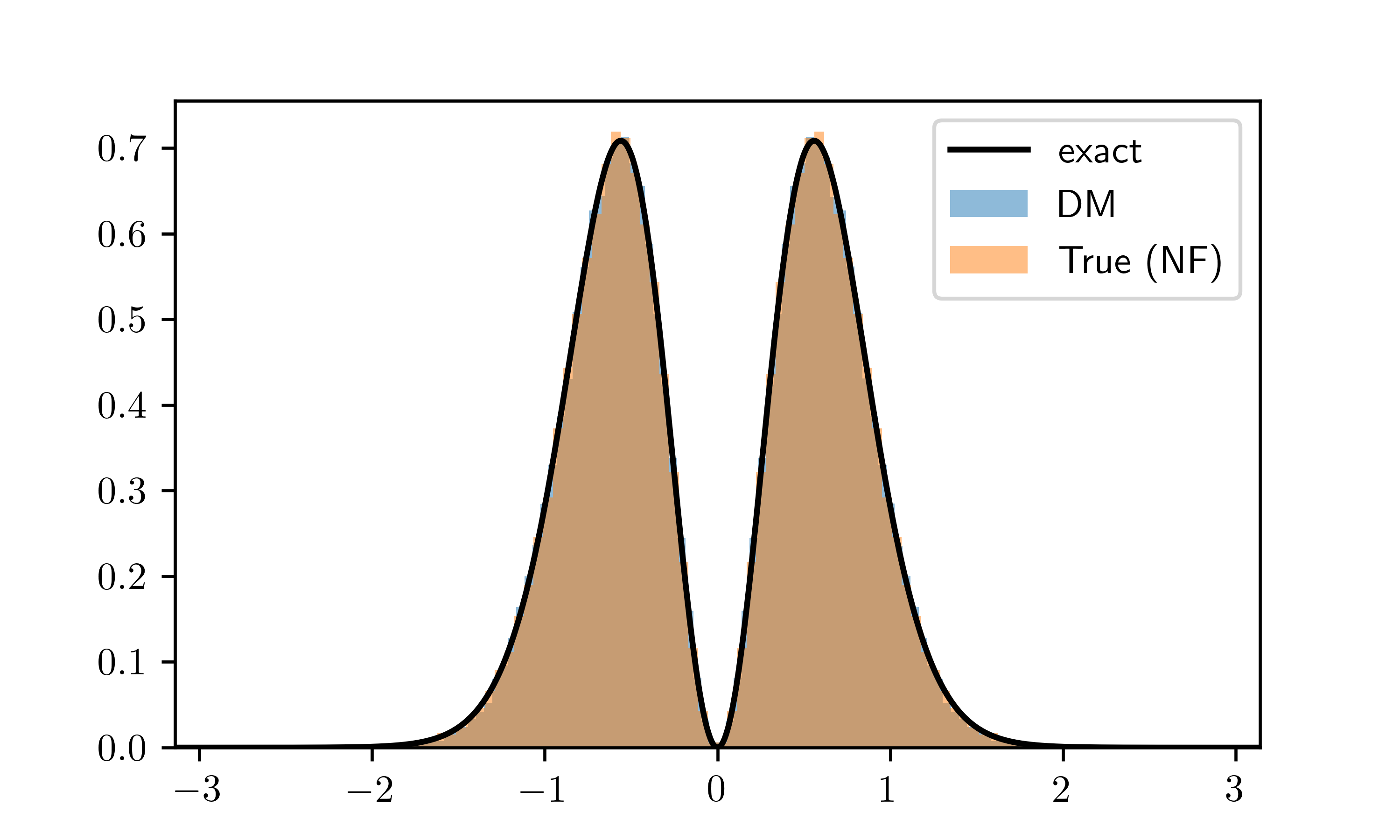}
    \includegraphics[width=0.48\linewidth, trim=0.5cm 0.2cm 1cm 0.5cm, clip]{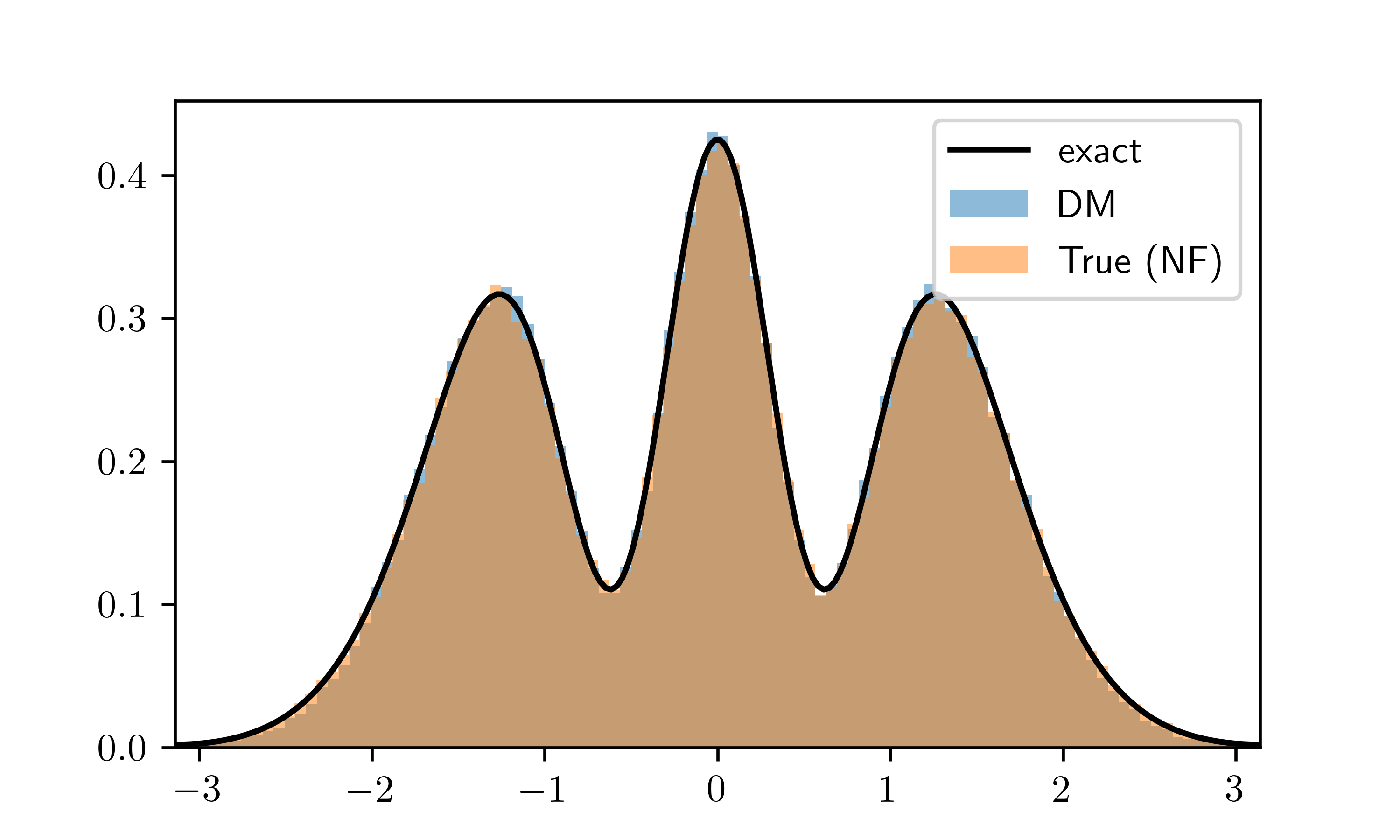}
    \caption{Application of diffusion-based sampling to SU(2) and SU(3) matrix models.
    Shown are the distributions of eigenangles obtained from the learned reverse diffusion process, compared with those from the training dataset and the exact PDFs. Left: SU(2). Right: SU(3).
    }
    \label{fig:SU_matrix_sampling}
\end{figure}

As a simple illustration, we apply the introduced score matching scheme to the SU(2) and SU(3) datasets whose exact and diffused eigenangles are shown in figure~\ref{fig:SU_matrix_eigval_diffusion}.
For the score function, we use a simple fully-connected neural network with explicit time dependence.
The network takes as input the real and imaginary parts of the group element together with the diffusion time,
and outputs a matrix in the Lie algebra $\mathfrak{su}(n)$.
Algebra constraints are enforced by projecting the network output onto the space of anti-Hermitian and traceless matrices.
Despite its simplicity, this model is sufficient to capture the structure of the target distribution in this toy setting.

We train the model for 600 epochs on our dataset, which includes 4096 samples for each group.
The trained score function is then used to generate samples from the diffusion process.
Figure~\ref{fig:SU_matrix_sampling} shows the distribution of eigen-angles of the generated samples alongside the training dataset.
The agreement is excellent, demonstrating that even a simple dense network can learn the score function in the toy setting.
\section{Application to SU(3) Gauge Theory}
\label{sec:gauge-theory}

In this section, we apply our score-matching diffusion framework to SU(3) lattice gauge theory,
with the goal of generating gauge field configurations distributed according to the Wilson gauge action:
\begin{align}
    p[U] &\propto \exp\left(-S_\text{W}[U]\right) \, ,\\
    S_\text{W}[U] &=
    -\frac{\beta}{2 N_c} \sum_{x \in \Lambda} \sum_{\mu \neq \nu}
    \mathbb{Tr}~ U_\mu(x) U_\nu(x + \hat \mu) U_\mu^\dagger (x + \hat\nu)U_\nu^\dagger (x)
    \,.
    \label{eq:Wilson:Plaq}
\end{align}
Here, $U_\mu(x)\in \mathrm{SU}(3)$ are the link variables, $\Lambda$ denotes the lattice sites,
$N_c=3$ is the number of colors, and $\beta$ is the inverse of gauge coupling.

To construct the generative model, we first introduce a gauge-equivariant score function that
respects the gauge symmetry of the Wilson action.
Next, we describe the methods that can be used for efficient sampling from learned score functions, including methods based on Langevin dynamics and HMD. We then present results from simulations in two dimensions, followed by results from simulations in four dimensions.

\subsection{Gauge-equivariant architecture for score function}
\label{sec:score-function}

The central object in the diffusion model is the time-dependent score function, which we parameterize using gauge-equivariant neural-network layers acting on the lattice link variables. In this work, we construct each layer from time-dependent combinations of Wilson \emph{staples}. These layers only mix nearby links, mirroring the locality of standard convolutional layers in machine learning. We refer to this building block as \texttt{GaugeLinkConv}, emphasizing that it is gauge-equivariant, defined directly on links, and convolution-like.

The structure of this gauge-equivariant layer, summarized in Algorithm~\ref{alg:gaugelinkconv}, is closely related to the L-CNN layers introduced in ref.~\cite{Favoni:2020reg}. The main difference is that L-CNN layers act on tuples consisting of nonlocally transforming gauge links together with locally transforming objects containing Wilson loops, whereas our construction operates only on the gauge links.

\begin{algorithm}[H]
\caption{\texttt{GaugeLinkConv} update for gauge link variables $U_\mu(x)\in \mathrm{SU}(3)$}
\label{alg:gaugelinkconv}
\begin{algorithmic}[1]
\Require Gauge link $U_\mu(x)$, time $t$, neighboring links needed to form staples

\State \textbf{Compute staples sum:}
\Statex $S_\mu(x)\gets \sum_{\nu\neq\mu}\Bigl(U_\nu(x+\hat\mu)\,U_\mu^\dagger(x+\hat\nu)\,U_\nu^\dagger(x)
+ 
U_\nu^\dagger(x-\hat\nu+\hat\mu)\,U_\mu^\dagger(x-\hat\nu)\,U_\nu(x-\hat\nu)\Bigr)$

\State \textbf{Time-dependent linear map (supporting multiple channels):}
\Statex $(s_1,s_2,s_3,s_4)\gets \bigl(w_1(t)S_\mu,\;w_2(t)S_\mu,\;w_3(t)S_\mu,\;w_4(t)S_\mu\bigr)$

\State \textbf{Gauge-equivariant update:}
\Statex $\tau_\mu(x)\gets \mathrm{Tr}\!\left(U_\mu(x)s_3 - s_4^\dagger U^\dagger_\mu(x)\right)$

\Statex $U_\mu(x)\gets \left(1+\frac{\tau_\mu(x)}{3}\right)U_\mu(x)$

\Statex $U_\mu(x)\gets U_\mu(x)+s_1^\dagger - U_\mu(x)\,s_2\, U_\mu(x)$

\State \textbf{Normalize using Frobenius norm:}
\Statex $U_\mu(x)\gets U_\mu(x)\big/ \left(\|U_\mu(x)\|/\sqrt{3}\right)$

\State \textbf{Return} $U_\mu(x)$
\end{algorithmic}
\end{algorithm}

\begin{figure}
    \centering
    \includegraphics[width=1\linewidth, trim=0 0 0 2cm, clip]{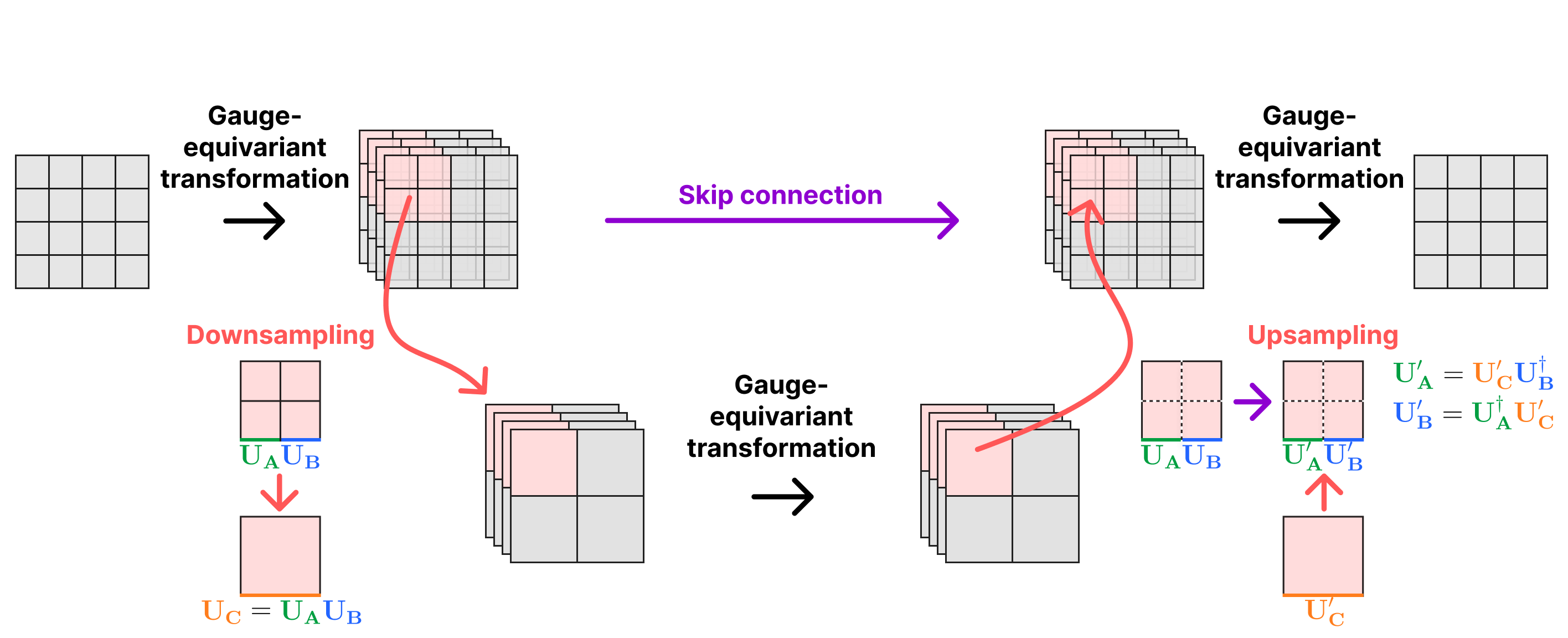}
    \caption{\texttt{GaugeLinkUNet} architecture in 2 dimensions with a single downsampling layer. The dashed links in the upsampling stage are directly taken from the corresponding skip connections.} 
    \label{fig:unet}
\end{figure}

Moreover, we construct a U-Net architecture~\cite{Ronneberger:2015unet}.
All operations within the U-Net respect gauge symmetry, and we therefore refer to it as \texttt{GaugeLinkUNet}. It consists of downsamplers, upsamplers, and a set of gauge-equivariant transformations such as \texttt{GaugeLinkConv} at each resolution level, as illustrated in figure~\ref{fig:unet} for the case of a single downsampling layer.

Downsampling is performed by multiplying neighboring gauge links with a fixed stride along each lattice direction.  
For example, along the direction \( \mu \), two adjacent links can be multiplied to form a coarse link:
\begin{equation}
    U^{(c)}_\mu(x) = U_\mu(x)\, U_\mu(x + \hat{\mu}).
\end{equation}

The upsampling stage reconstructs fine lattice representations from coarse ones while preserving gauge covariance.
Each coarse link \( U_\mu^{(c)}(x) \) is decomposed into finer links \( U_\mu(x) \) and \( U_\mu(x+\hat{\mu}) \) through 
\[
U_\mu(x) = U_\mu^{(c)}(x)\,U_{\mu,  \text{skip}}^\dagger(x+\hat{\mu}), 
\qquad 
U_\mu(x+\hat{\mu}) = U_{\mu, \text{skip}}^\dagger(x) \,U_\mu^{(c)}(x).
\]
Skip connections provide the fine-scale context \( U_{\mu, \text{skip}}(x) \) and \( U_{\mu, \text{skip}}(x+\hat{\mu}) \), taking directly from the fine lattice prior to downsampling.

The U-Net structure is expected to be useful for large lattices, where it can capture physics at multiple length scales. In this work, we mainly focus on small lattices and, therefore, do not expect a significant gain from using the U-Net structure.

\subsection{Methods for sampling via the reverse diffusion}

The reverse-time dynamics in eq.\eqref{eq:denoising:U_t} admits a \emph{family}
of generative processes, parameterized by a time-dependent function
$\tilde{\sigma}(t)$.
In general, the reverse-time dynamics can be solved as an SDE.
Alternatively, by setting $\tilde{\sigma}(t)=0$, the stochastic term vanishes and the
reverse process reduces to a deterministic ODE.
This deterministic formulation can be solved using conventional integrators that respect
the group structure.
For instance, one can use a fourth-order Runge--Kutta (RK4) method combined with a projection step that maps each update back onto SU(3).

To improve the accuracy of the reverse-time evolution, we also consider a
Predictor--Corrector (PC) framework~\cite{Song:2020Langevin}, summarized in
Algorithm~\ref{alg:pc_sun}. The predictor evolves the configuration backward in time
using either an SDE or ODE step, while the corrector refines the sample at fixed
diffusion time using a short Markov process.
Possible choices for the corrector include Langevin dynamics~\cite{Parisi:1980nh}
as well as Hamiltonian-based methods such as HMC (when applicable) and
HMD with randomized initial momenta.
Related approaches based solely on correction steps have also been explored in literature, including
annealed Langevin dynamics~\cite{Song:2019Langevin, Song:2020Langevin} and its
Metropolis-adjusted variant~\cite{Zhu:2025pmw}.

In the numerical simulations, we consider two methods for sampling.
The first method is based on integrating the reverse-time ODE using RK4 with projection onto SU(3) at each step.
The second method is PC sampling, in which the predictor is the same RK4-based ODE integrator,
while the corrector consists of short HMD trajectories evolved
with a leapfrog scheme; see Algorithm~\ref{alg:hmd_corrector_sun}.
For completeness, we also tested an alternative Langevin-based corrector, shown in Algorithm~\ref{alg:langevin_corrector_sun}.
The key difference between these correctors is the way stochasticity is introduced:
the Langevin corrector injects fresh Gaussian noise at every iteration,
whereas the HMD corrector samples the noise only once at the beginning (interpreted as an
initial momentum) and subsequently evolves the system deterministically.

\begin{algorithm}[t]
\caption{Predictor--Corrector Sampling}
\label{alg:pc_sun}
\begin{algorithmic}[1]
\State \textbf{Input:} Time grid $\{t_i\}_{i=0}^{N}$; initial state $U_0$;
score function $s_\theta(t, U)$. 
\State $U \gets U_0$
\For{$i = 0$ to $N-1$}
    \Comment{Predictor: advance $t_i \to t_{i+1}$}
    \State $U \gets \mathrm{Predictor}\!\left(s_\theta,\, t_i,\, t_{i+1},\, U\right)$
    \Comment{Corrector: refine at frozen $t_{i+1}$}
    \State $U \gets \mathrm{Corrector}\!\left(s_\theta,\, t_{i+1},\, U\right)$
\EndFor
\State $U \gets \mathrm{Predictor}\!\left(s_\theta,\, t_{N-1},\, t_{N},\, U\right)$  \Comment{Predictor only (no corrector at $t_N$)}
\State \Return $U$
\end{algorithmic}
\end{algorithm}

\begin{figure}[t]
\centering

\begin{minipage}[H]{0.48\textwidth}
\begin{algorithm}[H]
\caption{HMD-based corrector}
\label{alg:hmd_corrector_sun}

\begin{algorithmic}[1]

\State \textbf{Input:} Frozen time $t$, current state $U_t$,
number of steps $M$, step-size $\epsilon$

\State \colorbox{pink!50}{$\eta = \eta^a T^a,\quad \eta^a \sim \mathcal{N}(0,I)$}

\For{$j = 1$ to $M$}
    \State \colorbox{blue!15}{$U_t,\, \eta \gets \mathrm{ODE\_step}\!\left(
        \epsilon, \,  U_t, \, \eta, \,  s_\theta(t, U_t)
    \right)$}
\EndFor

\State \Return $U_t$

\end{algorithmic}
\end{algorithm}
\end{minipage}
\hfill
%
\begin{minipage}[H]{0.48\textwidth}
\begin{algorithm}[H]
\caption{Langevin-based corrector}
\label{alg:langevin_corrector_sun}

\begin{algorithmic}[1]

\State \textbf{Input:} Frozen time $t$, current state $U_t$,
number of steps $M$, noise scale $\alpha_t$

\For{$j = 1$ to $M$}
    \State \colorbox{pink!50}{$\eta = \eta^a T^a,\quad \eta^a \sim \mathcal{N}(0,I)$}
    \State \colorbox{blue!15}{$U_t \gets \exp\!\left(
        \alpha_t\, s_\theta(t, U_t)
        + \sqrt{2 \alpha_t}\, \eta
    \right)\, U_t$}
\EndFor

\State \Return $U_t$

\end{algorithmic}
\end{algorithm}
\end{minipage}

\end{figure}

To ensure exactness of the samples, one may enforce detailed balance through the Metropolis–Hastings procedure at the end of sampling,
for example via a few HMC updates. In the present work, however, we focus only on the performance of the unadjusted samplers, leaving accept–reject corrections to future work.
Consequently, the results presented below are biased, and no corrections are applied.

\subsection{Simulation results: 2D lattice}

In this subsection we present numerical results for SU(3) lattice gauge theory in two dimensions, focusing on the behavior of Wilson-loop observables along the forward diffusion process and the corresponding reverse-time sampling procedure.
Our main goal is to assess whether the learned reverse dynamics, initialized from the uniform distribution at $t=1$, can reproduce the target distribution at $t=0$.
We quantify performance for three different couplings and compare the reverse diffusion results against an HMC baseline started from the same initialization.

We apply the diffusion model to SU(3) gauge configurations generated with the Wilson action on a $16\times16$ lattice and inverse couplings $\beta = 4, 6, 12$.
For each value of $\beta$, we generate a training dataset of 4096 gauge-field configurations using HMC.
The score function is parameterized using four layers of \texttt{GaugeLinkConv}, in which the number of channels change as $1 \to 2 \to 4 \to 4 \to 1$.
In all three cases, the network is trained for 300 epochs using the implicit score-matching objective introduced above, as described in section~\ref{sec:score-matching}.

\begin{figure}
    \centering
    
    \includegraphics[width=1\linewidth, trim=3.5cm 8.4cm 3.5cm 1cm, clip]{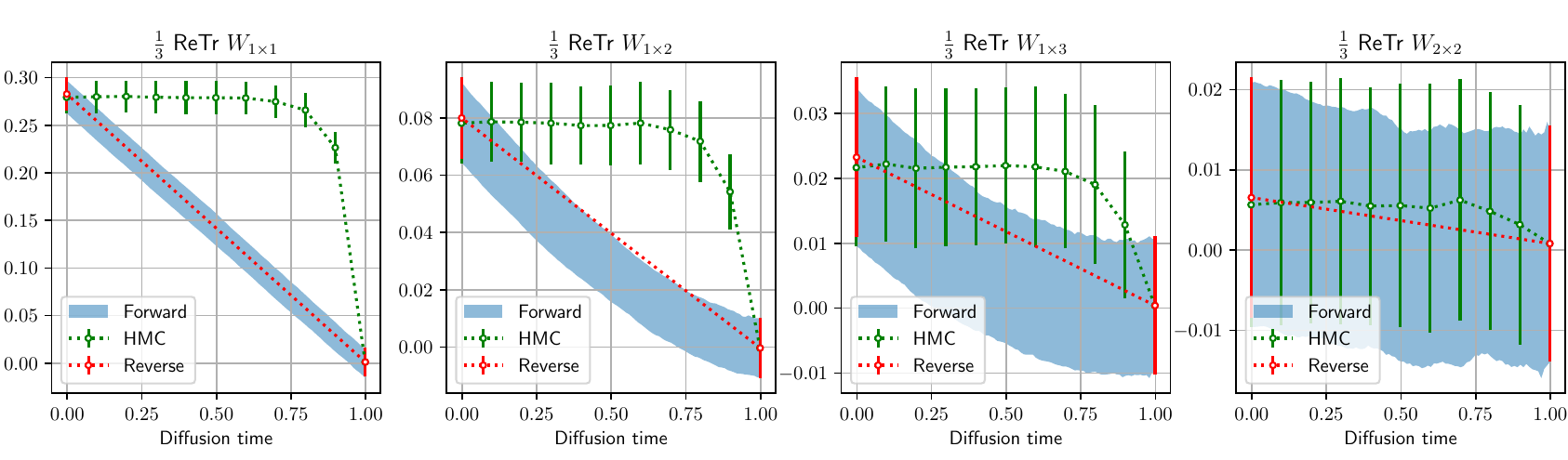}
    \\
    \includegraphics[width=1\linewidth, trim=3.5cm 8.4cm 3.5cm 1cm, clip]{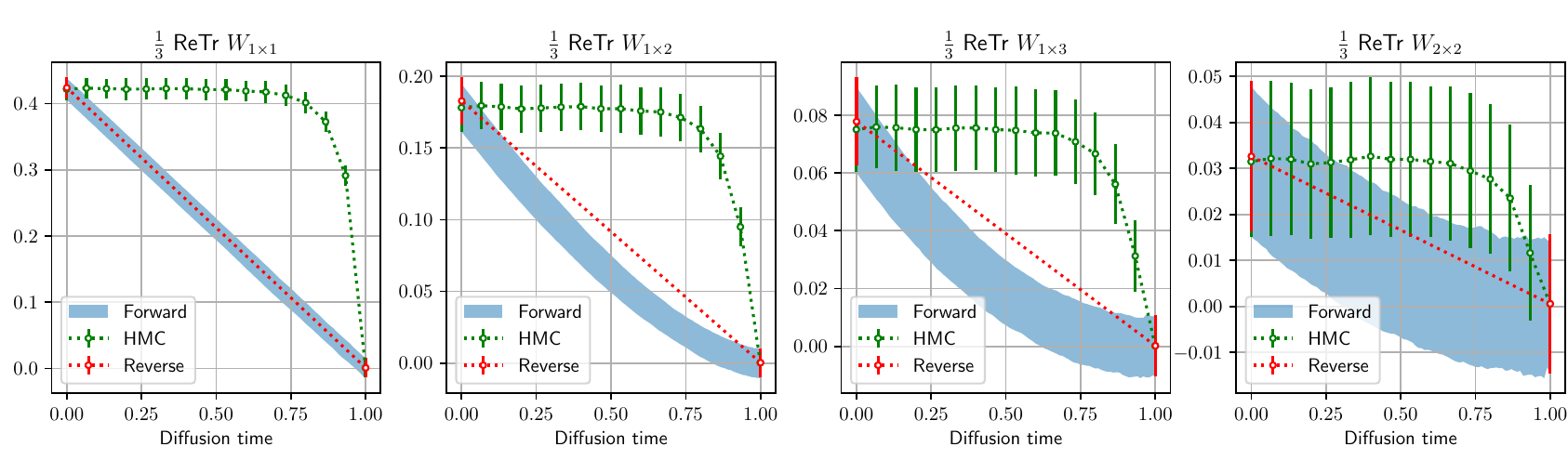}
    \\
    \includegraphics[width=1\linewidth, trim=3.5cm 8.4cm 3.5cm 1cm, clip]{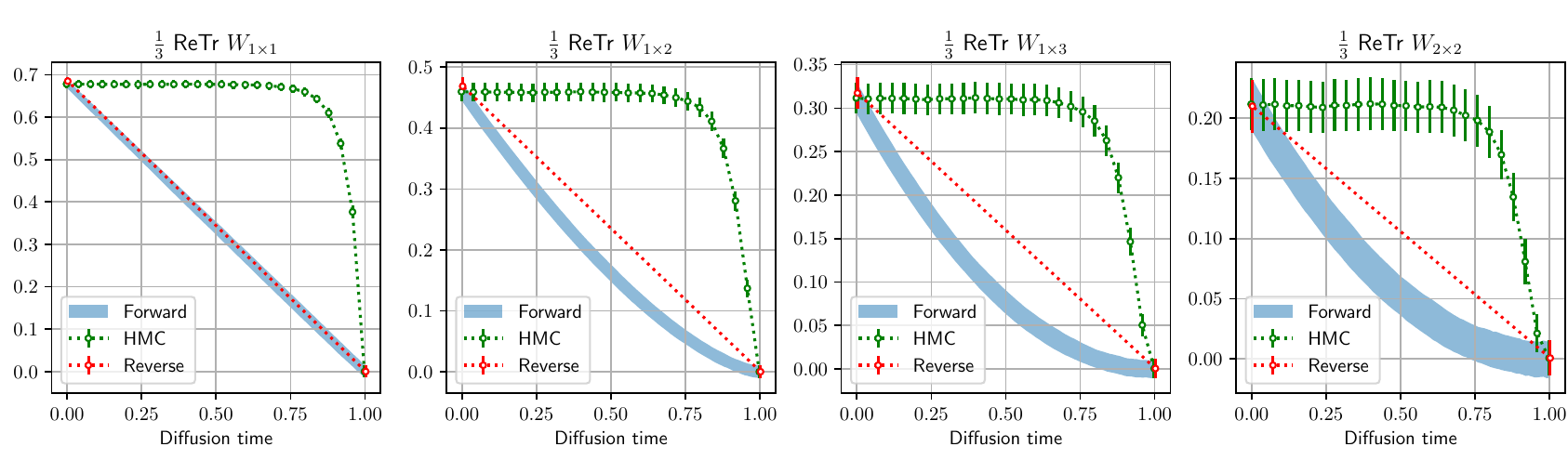}
    \caption{Application to SU(3) gauge theory on $16 \times 16$ lattices at $\beta=4$ (top), $\beta=6$ (middle), and $\beta=12$ (bottom).
    Columns (left to right) show the normalized trace of $1\times1$, $2\times1$, $3\times1$, and $2\times2$ Wilson loops.
    The horizontal axis is the diffusion time $t\in[0, 1]$.
    Blue bands: forward diffusion starting from true samples at $t=0$.
    Red symbols: reverse process starting from uniform samples at $t=1$.
    Green symbols: HMC trajectories initialized from the uniform distribution, plotted versus reversed diffusion time for comparison.
    Half-width of the bands and bars corresponds to one standard deviation over the corresponding ensemble of configurations.
    }
    \label{fig:SU3:gauge:16x16}
\end{figure}

Figure~\ref{fig:SU3:gauge:16x16} shows results on $16\times16$ lattices at $\beta=4$ (top row), $\beta=6$ (middle row) and $\beta=12$ (bottom row).
Columns from left to right display the normalized trace of the $1\times1$, $2\times1$, $3\times1$, and $2\times2$ Wilson loops as a function of the diffusion time $t\in[0,1]$,
measuring how well the learned model reproduces the target distribution at increasing
length scales.
The blue bands represent the forward diffusion process applied to a test dataset consisting of 1024 true configurations at $t=0$, generated using HMC.
As $t$ increases, the Wilson-loop traces evolve smoothly toward the uniform distribution at $t=1$, illustrating how the diffusion progressively erases structure from the gauge fields.
The half-width of the blue bands corresponds to one standard deviation over 1024 configurations.
We observe that different loop sizes decay at different rates: larger loops tend to be more sensitive to injected noise and approach their $t=1$
values more rapidly, while smaller loops evolve more linearly with $t$.

To evaluate the reverse-time generative procedure, we initialize 1024 configurations from the uniform distribution at $t=1$ and integrate the learned reverse-time dynamics toward $t=0$ by solving the ODE \eqref{eq:denoising:U_t:ODE} using a single RK4 step.
As the red bars in figure~\ref{fig:SU3:gauge:16x16} show, this choice already reproduces all four Wilson-loop observables accurately within one standard deviation.
As an additional baseline, we also show results from HMC trajectories initialized from the same uniform distribution, plotted against the reversed
diffusion time (green bars).
As a side remark, the last green bars coincide with the first points of the blue bands, since the equilibrated HMC configurations are used to construct the test dataset at $t=0$.

We also compare the computational cost of reverse diffusion sampling and HMC in terms of the number of staple computations.
Each leapfrog step in HMC requires computing the sum of all adjacent staples once.
The HMC trajectories use 10 to 15 leapfrog steps depending on $\beta$.
In the reverse diffusion sampler, a single RK4 step requires four score evaluations.
Each score evaluation involves collecting adjacent staples 11 times (sum of input channels), for a total of $4\times11 = 44$ staple computations per generated configuration.
Thus, a single RK4 step of our model is equivalent to 4.4 to 3 HMC trajectories in terms of staple computations in the settings considered here.

Thermalization of HMC simulations requires sufficiently many number of trajectories.
In order to compare the cost of generating samples using the learned score function with that of producing completely independent samples via HMC, one must fix a reference number of HMC trajectories.
As a practical reference point for the present study
and simply based on observables illustrated on figure~\ref{fig:SU3:gauge:16x16}, we assume that approximately four HMC trajectories suffice for $\beta=4$, whereas up to tens of trajectories may be required for $\beta=12$ to obtain approximately thermalized samples.
This assumptions are only for providing a reference for comparing computational costs.
Then, the overall cost of reverse diffusion sampling is comparable to HMC for small $\beta$, and becomes favorable as $\beta$ increases.
We note, however, that this heuristic comparison does not account for additional overhead or exactness corrections.

Unlike the DM-based samples generated here, HMC provides exact samples.
In principle, diffusion-based sampling can also be made exact by incorporating explicit corrections, such as a Metropolis–Hastings accept–reject step or an importance-sampling reweighting.
We do not study such corrections here, leaving the construction of an exact diffusion-based sampler for future work.

Finally, we emphasize that here we restrict our analysis to four Wilson-loop observables. In principle, one should also examine larger Wilson loops and more global quantities, such as the topological charge, which probe longer-range structure of the gauge fields. From such observables one may find that a purely ODE-based sampler is not always sufficient to produce high-quality samples. In those situations, a predictor–corrector framework can be employed to improve sampling accuracy. Although such a strategy could also be applied in the present two-dimensional setting, we do not pursue it here, since its role becomes more significant in four dimensions. We therefore defer the use of predictor–corrector sampling to the next subsection, where we study the four-dimensional theory.

\subsection{Simulation results: 4D lattice}

We now consider SU(3) gauge theory in four dimensions, where the Wilson action couples a larger number of plaquettes per link.
This makes the target distribution more challenging for generative modeling and reverse-time integration.
For this reason, we restrict our study to smaller inverse couplings, focusing on $\beta=3$ and $\beta=6$ on a $4^4$ lattice, in contrast to the two-dimensional results at $\beta=4,6,12$ presented above.

For each value of $\beta$, we generate a training dataset of 4096 gauge-field configurations using HMC.
For $\beta=3$, the score function is parameterized using four layers of \texttt{GaugeLinkConv}, in which the number of channels change as $1 \to 4 \to 4 \to 4\to1$.
For $\beta=6$, we use a six-layer architecture with $1 \to 4 \to 4 \to 4 \to 4 \to 4 \to 1$ channels.
In both cases, the network is trained for 300 epochs using the implicit score-matching objective introduced above.

\begin{figure}
    \centering
    \includegraphics[width=1\linewidth, trim=3.5cm 8.4cm 3.5cm 1cm, clip]{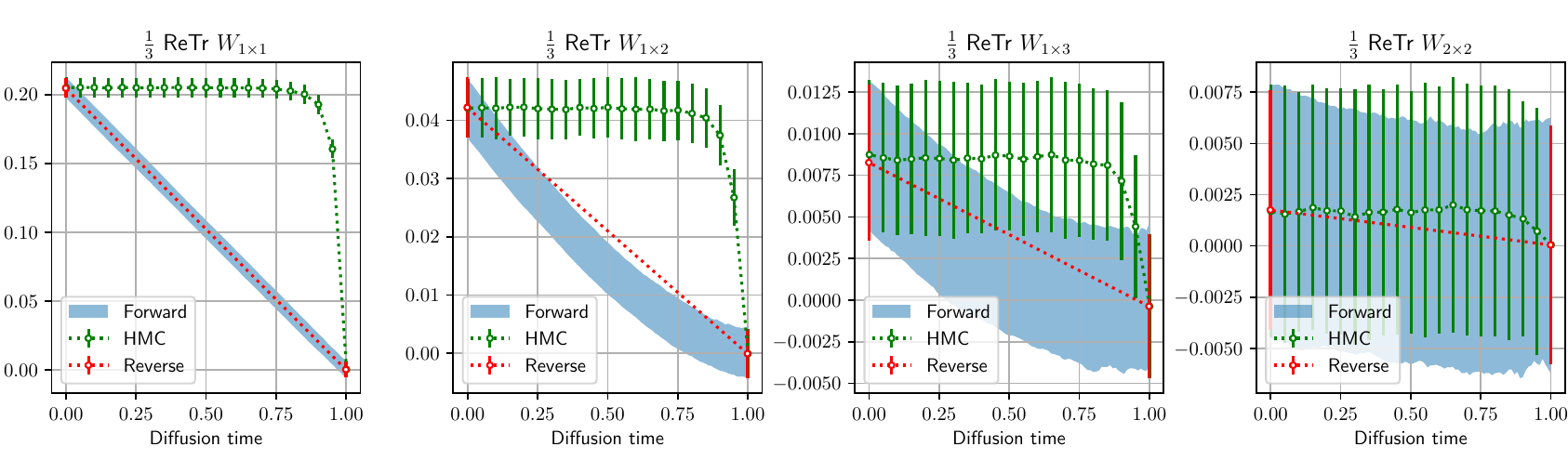}
    \\
    \includegraphics[width=1\linewidth, trim=3.5cm 8.4cm 3.5cm 1cm, clip]{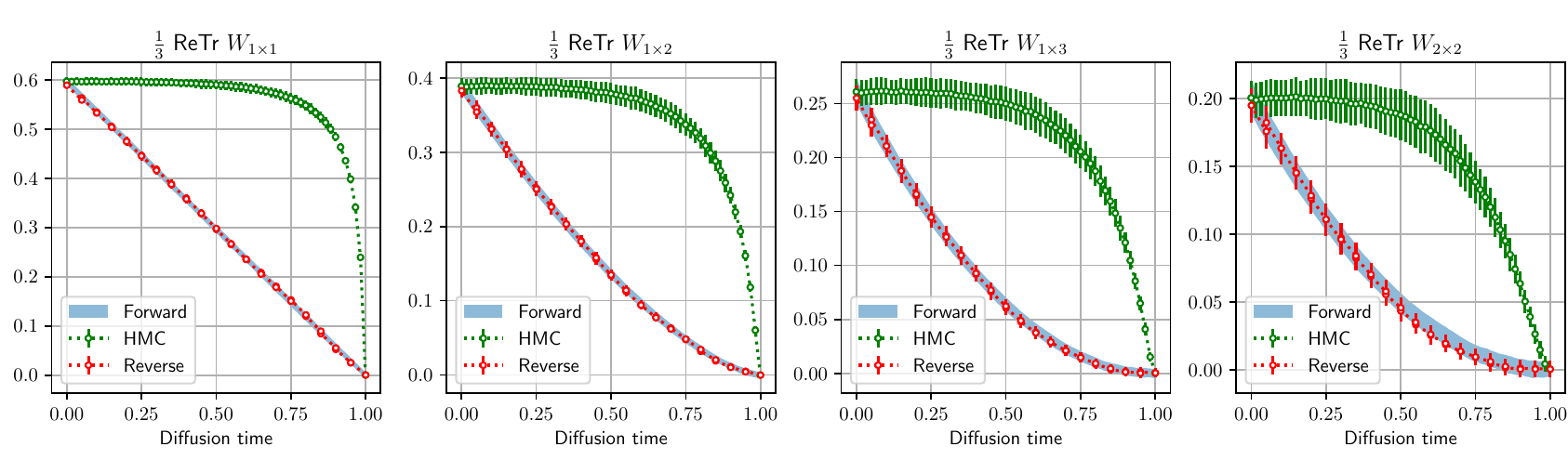}
    \caption{
    Application to SU(3) gauge theory on $4^4$ lattices at $\beta=3$ (top) and $\beta=6$ (bottom).
    Plot conventions are the same as in figure~\ref{fig:SU3:gauge:16x16}.
    }
    \label{fig:SU3:gauge:4x4x4x4}
\end{figure}

Figure~\ref{fig:SU3:gauge:4x4x4x4} shows results on a $4^4$ lattice at $\beta=3$ (top) and $\beta=6$ (bottom).
As in the 2D studies, we track Wilson-loop observables as a function of the diffusion time $t\in[0,1]$.
The blue bands represent the forward diffusion process applied to 1024 (top) and 256 (bottom) true configurations at $t=0$, illustrating how the observables evolve as noise is injected and the distributions approach the uniform distribution at $t=1$.
We also plot the Wilson-loop observables obtained from HMC trajectories used to generate the correct samples,
shown against the reversed diffusion time (green bars).

The numerical scheme required for reliable sampling depends on the inverse coupling.
For $\beta=3$, we reverse the process by solving the ODE \eqref{eq:denoising:U_t:ODE} using a single RK4 step.
As shown in the top panels of figure~\ref{fig:SU3:gauge:16x16}, this choice already reproduces all four Wilson-loop observables at $t=0$ within one standard deviation.

For $\beta=6$, however, a pure ODE-based method is not sufficient.
In this case we employ a predictor--corrector scheme: the predictor integrates the ODE \eqref{eq:denoising:U_t:ODE} using 20 RK4 steps, and after each predictor step (except the final one) we apply an HMD-based corrector.
The red bars in the lower panels of figure~\ref{fig:SU3:gauge:16x16} also show the intermediate values produced by the predictor--corrector procedure; the predictor and corrector values differ by less than one standard deviation.

We now compare the computational cost of reverse diffusion sampling and HMC in terms of the number of staple computations.
For $\beta=3$, we use a four-layer architecture, and a single RK4 step is roughly equivalent to 5 HMC trajectories (with 10 leapfrog steps) in terms of staple computations.
Analogous to the 2D experiments, we compare the cost of generating samples using the learned score function with that of producing fully independent samples via HMC, by fixing a reference number for HMC trajectories.
For the purpose of comparison, we heuristically assume that HMC thermalization requires approximately 10 trajectories.
Therefore, in this regime the overall cost of reverse diffusion sampling is about one half of the HMC thermalization cost, ignoring additional overhead and the absence of exactness corrections.

At $\beta=6$, the problem becomes significantly more challenging for reverse diffusion, and to some extent also for HMC (see the lower panels of
figure~\ref{fig:SU3:gauge:4x4x4x4}).
In the figure we use 60 HMC trajectories (with 15 leapfrog steps each), while the reverse diffusion sampler employs 20 RK4 steps together with an HMD-based corrector after every step except the last one.
In terms of the number of staple computations, each RK4 step with the six-layer architecture is roughly equivalent to $4 \times 21 / 15$ HMC trajectories.
This implies that the cost of 20 RK4 steps alone is already roughly equivalent to 112 HMC trajectories.
Including the cost of the HMD corrections further increases the total cost of reverse diffusion sampling to about 511 HMC trajectories, which is several times larger than the HMC thermalization cost.

\begin{figure}
    \centering
    \includegraphics[width=1\linewidth, trim=3.5cm 8.4cm 3.5cm 1cm, clip]{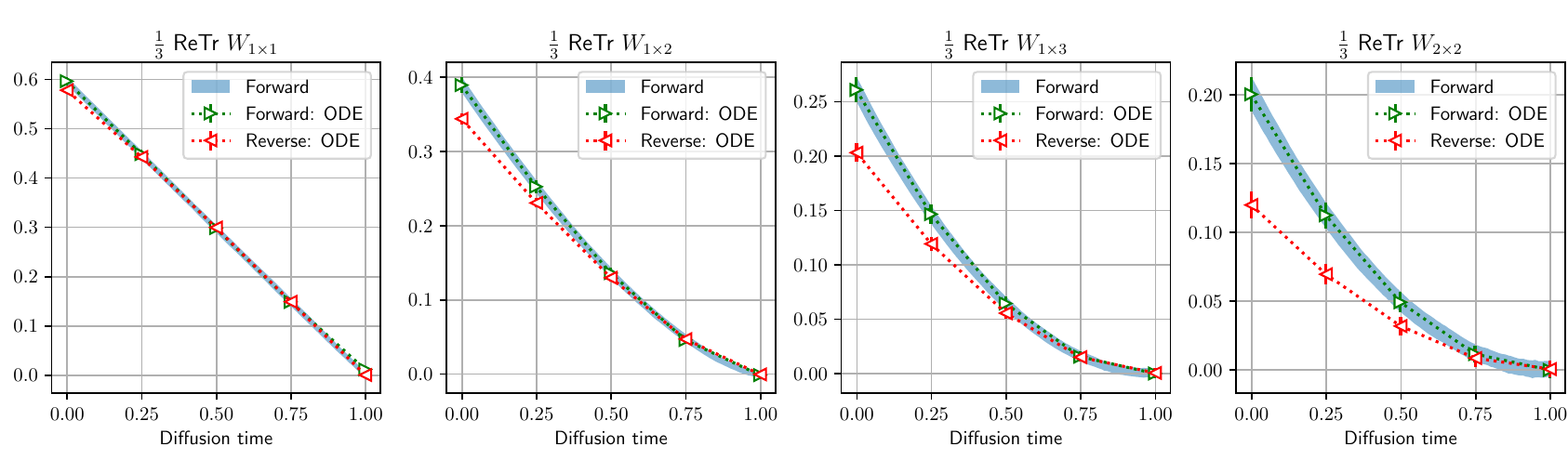}
    \caption{
    Forward vs reverse dynamics of SU(3) gauge theory on a $4^4$ lattice at $\beta=6$.
    Plot conventions follow figure~\ref{fig:SU3:gauge:16x16},
    except for the green symbols, which here represent the forward ODE solution initialized from equilibrium configurations.
    }
    \label{fig:SU3:gauge:4x4x4x4:forward-vs-reverse}
\end{figure}

We find an ODE-based method without correctors insufficient for sampling the theory at $\beta=6$.
We now explain why this happens.
Figure~\ref{fig:SU3:gauge:4x4x4x4:forward-vs-reverse} illustrates the forward and reverse dynamics for SU(3) gauge theory on a $4^4$ lattice at this inverse coupling.  
The green symbols show the forward ODE evolution starting from equilibrium configurations at $t=0$ using four RK4 steps, demonstrating that the learned score function closely follows the stochastic forward diffusion process, with only a small deviation near $t=1$.  
By contrast, the reverse-time ODE evolution, initialized from uniform configurations at $t=1$, approximately retraces the diffusion process but gradually deviates as $t$ decreases.  
This deviation is much more pronounced than for the forward ODE: the forward evolution tracks the stochastic diffusion closely up to $t=1$, with differences hardly visible, whereas the reverse ODE lags behind the target by up to roughly 40\% for the $2\times 2$ Wilson loop in the far-right panel.  
This behavior persists even with smaller ODE step sizes, indicating that it is not caused by discretization errors.

The deviation of the reverse-time ODE arises primarily from a mismatch in the initial distribution at $t=1$.  
The diffusion process, with the noise schedule in eq.~\eqref{eq:noise:schedule}, guarantees that the distribution at $t=1$ is exactly uniform.
However, when the forward ODE is solved using the learned score function, it produces a distribution that is close to, but not exactly, uniform.  
Consequently, the reverse ODE sampler can yield accurate results only when initialized from the same distribution as the terminal distribution of the forward ODE evolution.  
The reverse ODE is sensitive%
\footnote{For a discussion of instabilities in ODE solutions, see refs.~\cite{Bender:2014rqx, Komijani:2021xya}.}
to the initial distribution, and small inaccuracies in the learned score function accumulate during the reverse evolution, causing the reconstructed Wilson loops to lag behind their target values.  
This effect becomes more pronounced at larger $\beta$, 
rendering a purely deterministic ODE insufficient.
In such cases, stochastic correction schemes, such as predictor--corrector methods with HMD or Langevin updates, can restore accuracy in reverse-time sampling.
Another direction to improve the sampling is to explicitly account for the mismatch at $t=1$ by learning a modified terminal distribution, for example through an additional diffusion model, so that reverse-time sampling can recover the correct equilibrium distribution without the need for costly correction steps.
\section{Summary and concluding remarks}
\label{sec:conclusion}

In this work, we developed a diffusion-based framework for sampling non-Abelian lattice gauge theories and applied it to SU(3) gauge theory in two and four dimensions. Our primary objective was to assess how diffusion-based sampling compares with the established HMC method in terms of computational cost and sampling efficiency.

In two dimensions on a $16\times16$ lattice, reverse diffusion successfully reconstructs several Wilson-loop observables starting from uniform configurations at $t=1$. Even a single RK4 step in the reverse-time ODE can already produce estimates that agree within one standard deviation; however, accurate and reliable sampling requires restoring exactness and may need a predictor--corrector scheme at large values of inverse coupling.

To assess computational efficiency, we compared diffusion-based sampling and HMC using the number of staple evaluations as a proxy. Within the setting of our experiments on the $16\times16$ lattice, we argued that diffusion-based sampling can achieve computational costs comparable to HMC or less as $\beta$ increases.

In four dimensions on a $4^4$ lattice, the computational costs can be substantially higher. While small inverse couplings resemble the 2D case, physically relevant regimes (e.g., $\beta=6$) require many reverse-time steps together with correction updates, making diffusion-based sampling several times more expensive than the HMC thermalization. Thus, more efficient gauge-equivariant architectures and improved reverse-time integration strategies are needed.

Several directions could reduce the computational cost. Architectural improvements and optimized implementations could lower the expense of score evaluation, particularly in four dimensions. In particular, more efficient equivariant networks, such as the U-Net architecture introduced here, may improve scalability. In our $4^4$ lattices, the possible benefits of the U-Net are not realized, but preliminary tests already suggest modest improvements in estimation of larger Wilson loops, e.g., $2\times3$, $2\times4$, and $2\times4$ Wilson loops.
The benefits of U-Net architecture are expected to be evident on large lattice volumes, where it can capture long-range correlations more efficiently.

The performance of the reverse diffusion process also depends on the choice of noise schedule. The schedule defined in eq.~\eqref{eq:noise:schedule} produces an approximately linear decay of the average plaquette toward zero, resulting in a smooth interpolation between $t=0$ and $t=1$. This smooth behavior allows accurate integration of the reverse diffusion process with only a small number of steps; in many experiments, even a single RK4 step sufficed. By contrast, a commonly used exponential schedule leads to an exponential decay of the average plaquette. This behavior is not inherently problematic, as its effects can be compensated by an appropriate change of variables during the reverse diffusion.


Another practical consideration concerns the corrector strategy. In this work, we introduced and employed an HMD-based corrector. We also tested an alternative Langevin-based corrector~\ref{alg:langevin_corrector_sun}. The corrector step size $\alpha_t$ is chosen according to a fixed signal-to-noise ratio (SNR) to balance the deterministic score update and stochastic noise. Specifically, following~\cite{Song:2020Langevin}, we set $\alpha_t = 2 (r_\mathrm{SNR}\, \| \eta \| / \| \mathbf{s} \|)^2$,
where $\mathbf{s} = \mathbf{s}_\theta(t, U_t)$ denotes the score estimate and $\eta$ is Gaussian noise in the Lie algebra. 
For $r_\mathrm{SNR} = 0.1$, we observed that the HMD-based corrector performs better in our experiments. We note, however, that this observation cannot be generalized, as fine-tuning of $\alpha_t$ can affect the efficiency of the Langevin-based corrector. Similarly, one can fine tune the parameters of the HMD corrector, such as its trajectory length and step size to optimize the importance.

Finally, extending this framework to larger lattices, a broader set of observables, and ultimately to gauge theories with dynamical fermions is an important direction for future research. Beyond architectural improvements, optimized corrector strategies will be essential for improving computational efficiency in large-scale applications. Exploring these directions may lead to efficient diffusion-based sampling methods for lattice gauge theory.

\acknowledgments

The authors are thankful to Gert Aarts, Aida El-Khadra, Tej Kanwar, Octavio Vega, and Lingxiao Wang for insightful discussions.
JK and MKM acknowledge the support received from the Horizon Europe project interTwin, funded by the European Union Grant Agreement Number 101058386.
MKM thanks the Kavli Institute for Theoretical Physics (KITP) for hospitality and support during the programs “What is Particle Theory?” and “Generative AI for High \& Low Energy Physics”. 
The KITP is supported in part by the National Science Foundation under Grant No. PHY-2309135. MKM is grateful for the hospitality of Perimeter Institute where part of this work was carried out. Research at Perimeter Institute is supported in part by the Government of Canada through the Department of
Innovation, Science and Economic Development and by the Province of Ontario through the Ministry of Colleges and Universities. This research was also supported in part by the Simons Foundation through the Simons Foundation Emmy Noether Fellows Program at Perimeter Institute.

\section*{Code Availability}
The primary codes used in this work are publicly available in the
\href{https://github.com/jkomijani/lattice_ml/tree/main/src/lattice_ml/diffusion}{diffusion},
\href{https://github.com/jkomijani/lattice_ml/tree/main/src/lattice_ml/gauge_tools}{gauge\_tools},
and
\href{https://github.com/jkomijani/lattice_ml/tree/main/src/lattice_ml/integrate}{integrate}
subpackages in 
\url{https://github.com/jkomijani/lattice_ml}.
Two notebooks for the SU(2) and SU(3) toy models are available in \href{https://github.com/jkomijani/lattice_ml/tree/main/examples/diffusion/matrix_models}{examples/diffusion/matrix\_models}.
Uniformly distributed SU$(N)$ link generation is implemented using the
\href{https://github.com/jkomijani/normflow/tree/main/src/normflow/prior}{prior} subpackage in
\url{https://github.com/jkomijani/normflow}.
The gauge-equivariant U-Net developed for this work is currently not publicly released; interested readers may contact the authors for access.

\appendix

\section{Mathematical Formalism}
\label{apx:formalism}

Let $\{T^a\}$ be the anti-Hermitian generators of the Lie algebra, normalized according to
\begin{equation}
    \Tr \left(T^a T^b\right) = - \delta^{ab}\, .
\end{equation}
Any element $A$ of the Lie algebra can then be expanded as
\begin{equation}
    A = A^a T^a, \qquad A^a = - \Tr \left(T^a A\right)\,,
\end{equation}
where $A_a$ are real.
For any arbitrary matrix $M$, we have
\begin{equation}
    \ReTr \left(T^a M\right) T^a = - \mathcal{P}_{\mathfrak g}(M)\,,
\end{equation}
where $\mathcal{P}_{\mathfrak{g}}$ denotes the projection operator onto the Lie algebra $\mathfrak{g}$.
The left derivative of a scalar function $f: G \to \mathbb{R}$ is defined by
\begin{equation}
    \partial^a f(U) \;\equiv\;
    \left. \frac{d}{ds}\, f\left(e^{s T^a} U \right)\right|_{s=0}.
\end{equation}
This definition can be naturally extended to complex-valued functions.
Expanding $f$ around $U$ gives the formal Taylor series~\cite{Drummond:1982sk}
\begin{equation}
    f\!\left(e^{\alpha^a T^a} U \right)
    = \left(1 + \alpha^a \partial^a + \frac{1}{2} \alpha^a \alpha^b \partial^a \partial^b + \cdots\right) f(U) \,.
    \label{eq:Taylor-expansion}
\end{equation}
Note that $\alpha_a$ may be a function of $U$; then, the ordering of multiplicative terms in the above expression matters because the derivatives act only on $f$ in the above expansion.
When $\alpha_a$ are constant, the above relation can be written
as
\begin{equation*}
    f\!\left(e^{\alpha^a T^a} U \right)
    = e^{\alpha^a \partial^a} f(U).
\end{equation*}

Integration over the group is performed using the Haar measure $dU$, which is invariant under
both left and right multiplication: $dU = d(V_0 U) = d (U W_0)$.
This invariance implies
\begin{equation}
    \int dU f\!\left(e^{\alpha^a T^a} U \right)
    = \int dU f(U) 
    \quad \Rightarrow\quad \int dU \partial^a f(U) = 0 \,.
\end{equation}
Consequently, we obtain the integration by parts rule
\begin{align}
    \int dU f(U)\, \partial^a g(U) = - \int dU g(U)\, \partial^a f(U)\,.
    \label{eq:int-part-rule}
\end{align}
A convenient exponential form of this relation is
\begin{align}
    \int dU f\!\left(e^{\alpha^a T^a} U\right) g(U)
    = \int dU f(U) \left(1 - \partial^a \alpha^a + \frac{1}{2} \partial^a \partial^b \alpha^a \alpha^b + \cdots\right) g(U)\,,
    \label{eq:int-part-rule:exp}
\end{align}
where the ordering on the right-hand side matters, since $\alpha_a$ may depend on $U$.

The delta distribution on the group is defined formally by
\begin{align}
\left\{
\begin{aligned}
    \int dU\, f(U)\, \delta_G(UV^\dagger - I) &= f(V)\,, \\
    \int dV\, f(V)\, \delta_G(UV^\dagger - I) &= f(U)\,.
\end{aligned}
\right.
\label{eq:delta-group}
\end{align}
Using this definition together with Eq.~\eqref{eq:int-part-rule:exp}, we obtain
\begin{align}
\left\{
\begin{aligned}
    \int dU f(U)\, \delta_G\!\left(e^{\alpha^a T^a} U V^\dagger - I\right)
    &= \left(1 - \partial^a \alpha^a + \frac{1}{2} \partial^a \partial^b \alpha^a \alpha^b + \cdots\right) f(V)\,,
    \\
    \int dV f(V)\, \delta_G\!\left(U V^\dagger e^{-\alpha^a T^a} - I\right)
    &= \left(1 - \partial^a \alpha^a + \frac{1}{2} \partial^a \partial^b \alpha^a \alpha^b + \cdots\right) f(U)\,.
    \displaystyle
\end{aligned}
\right.
\label{eq:int-part-rule:exp:delta}
\end{align}
It is evident from the context that the derivative operator is taken with respect to $V$ and $U$ on the right hand side of the first and second lines, respectively.


We now present the formal Fourier representation of the delta distribution on a compact Lie group. 
Exploiting an exponential representation of the group, the group delta function can be expressed as an integral over the Lie algebra:
\begin{align}
   \delta_{G}(U - I) &= \int dA\, e^{i\, \ReTr \left[\log(U) A\right]},
   \label{eq:delta_G:Fourier}
\end{align}
where $A = A_a T^a$ is an element of the Lie algebra, 
and $dA = c_0 \prod_a dA_a$ denotes the invariant measure on the algebra, with $c_0$ fixing the normalization.
The real part emphasizes the integrand is well defined for complex-valued matrices
although in principle the trace returns a real value.
The logarithm has multiple values corresponding to different images of the identity.

To compute derivatives of the group delta function, one can use the integral representation of the matrix logarithm,
\begin{equation}
    \log(U) = (U - I) \int_0^1 \frac{ds}{I + s(U - I)}\,,
\end{equation}
which is valid for matrices without negative real or zero eigenvalues.
If $U$ has an eigenvalue on the negative real axis, the integration contour must be slightly deformed to avoid the singularity.
This expression allows for algorithmic matrix differentiation:
\begin{align}
  \partial^a \log(U) &= \int_0^1 ds\,
  \big[I + s(U - I)\big]^{-1} T^a U \big[I + s(U - I)\big]^{-1} .
\end{align}
Although this form looks complicated, it simplifies when
evaluated at $U = I$, where the group delta function is supported. 
Expanding about the identity yields
\begin{equation}
   \partial^a \log(U)\Big|_{U = I} = T^a .
\end{equation}
Another practical case is
\begin{equation}
  \Tr \left[ f(U)\, \partial^a \log(U) \right]
  = \Tr \left[f(U) T^a\right],
\end{equation}
which yields the following useful formula
\begin{equation}
  T^a \partial^a e^{-\frac{1}{2} \ReTr \left[ \log(U) \log(U^\dagger) \right]}
  = - \log(U)\, e^{-\frac{1}{2} \ReTr \left[ \log(U) \log(U^\dagger) \right]} , \label{eq:score:group-Guassian}
\end{equation}
where we used the fact that $\log(U)$ is anti-Hermitian.
Note that there is a minus sign due to the normalization of the generators that cancels the minus from $\log(U^\dagger)$.

We now return to the discussion about the group delta function.
A useful identity involving the derivative of the group delta function follows directly from 
the Fourier representation \eqref{eq:delta_G:Fourier}. 
Applying the left-invariant derivative to $\delta_G(U - I)$ gives
\begin{align}
    T^a \partial^a \delta_{G}(U - I)
    &= T^a \int dA\,\left. \frac{d}{d\epsilon}
    e^{i\, \ReTr \left[\log\left(e^{\epsilon T^a} U\right) A\right]}
    \right|_{\epsilon=0} \nonumber \\
    &= i\, T^a \int dA\,\ReTr [T^a A]\, e^{i\, \ReTr \left[\log(U) A\right]} \, \nonumber \\
    &= -i\, \int dA\, A \, e^{i\, \ReTr \left[\log(U) A\right]}\,.
    \label{eq:co-deriv-of-exp}
\end{align}
where the minus sign in the last line follows again from the normalization of the generators.
\section{Diffusion process on SU($N$) matrices}
\label{apx:Diffusion-process}

\subsection{Diffusion process and Fokker-Planck equation}

We consider a diffusion process acting on SU($N$) gauge link variables on a lattice.
The following discussion can be also extended to other Lie groups.
For each lattice link variable located at site $x$ in direction $\mu$,
we denote its value at (fictitious) diffusion time $t$ as
\begin{equation}
    U_t \equiv U(t, x, \mu)\,.
\end{equation}
To simplify notation, we suppress explicit dependence on $(x, \mu)$.

The evolution of $U_t$ is defined recursively by the stochastic update rule
\begin{equation}
   U_{t + h} = \exp\!\left(h f(t, U_t) + \sqrt{h} \sigma \eta(t)\right) U_t,
   \label{eq:diffusion:recursive:U_t}
\end{equation}
where $h$ is an infinitesimal time step and the process approaches
a continuous-time stochastic differential equation (SDE) as $h \to 0$.
Here,
\begin{equation}
    \eta(t) = \eta^a(t)\, T^a\,,
\end{equation}
is a complex Gaussian random matrix with components $\eta_a(t)$ drawn independently
from a normal distribution of zero mean and unit variance.
The matrices $\{T^a\}$ are generators of the gauge group Lie algebra; see
Appendix~\ref{apx:formalism} for our notations and conventions.
Both the deterministic drift $f(t, U_t)$ and the noise term $\eta(t)$ are elements of the algebra.
The real parameter $\sigma$ controls the noise amplitude and may, in general, depend on the lattice
position and direction $(x, \mu)$ and on time $t$, although for compactness of notation we omit these indices.

To derive the corresponding Fokker–Planck (FP) equation governing the evolution of the probability density $\rho_t(U)$,
we express the density at time $t + h$ in terms of $\rho_t(U_t)$:
\begin{align}
    \rho_{t + h}(U)
    &= \mathbb{E}_{\eta}\!\left[
    \int dU_t\, \rho_t(U_t)\,
    \delta_G\!\left( U U_t^\dagger e^{-h f(t, U_t) - \sqrt{h} \sigma \eta} - I \right)
    \right].
    \label{eq:rho:recursive}
\end{align}
One can use Eq.~\eqref{eq:int-part-rule:exp:delta}
to evaluate the integral.
Then, keeping terms of order $h$, and calculating the expectation values over $\eta$,
we obtain
\begin{align}
    \rho_{t + h}(U)
    &= \Big(1 - h\, \partial^a f^a(t, U) + \frac{h \sigma^2}{2} \partial^a \partial^a
    \Big)\, \rho_t(U)
    + \mathcal{O}(h^2)\,.
\end{align}
Taking the limit $h \to 0$, we obtain the continuous-time form of the FP equation:
\begin{align}
    \partial_t \rho_t(U)
    &= \partial^a \!\left(- f^a(t, U) + \frac{\sigma^2}{2}\, \partial^a \right)
    \rho_t(U)\,.
    \label{eq:FP:Lie}
\end{align}
Higher-order terms in $h$ can be systematically included to capture discretization effects if higher accuracy is required.

Equation~\eqref{eq:FP:Lie} describes the time evolution of the probability distribution of the gauge link variables.
The first term represents the deterministic drift and the second term corresponds to stochastic diffusion induced by the Brownian motion on
SU($N$).
The score function, i.e., the gradient of the log-probability with respect to
the gauge field $U$,
\begin{equation}
    \mathbf{s}(t, U) = T^a \partial^a \log \rho_t(U)
\end{equation}
appears in the FP equation and plays a critical role in guiding the reverse diffusion process.

\subsection{Reverse Process}

The process defined in Eq.~\eqref{eq:diffusion:recursive:U_t} is reversible in the sense that evolving the random field $U_{t_1}$ forward in time from $t_1$ to $t_2$
using Eq.~\eqref{eq:diffusion:recursive:U_t}, and subsequently evolving it backward from 
$t_2$ to $t_1$, leaves the probability density function (PDF) of $U_{t_1}$ unchanged. Equivalently, the forward- and reverse-time processes must satisfy the same Fokker–Planck equation.

As in the Euclidean setting discussed in Appendix B of \cite{Vega:2025hgz}, the reverse-time evolution on the group manifold is not unique. Instead, it forms a family of valid reverse processes, parameterized by an effective noise amplitude $\tilde \sigma$, which may, in general, depend on the lattice position and direction $(x, \mu)$,
but not on the gauge link variables.

Following the arguments of Appendix B in \cite{Vega:2025hgz}, the reverse-time stochastic update can be expressed as
\begin{equation}
   U_{t - h}
   = \exp\!\left(
       -h\, f(t, U_t)
       + \frac{ (\sigma^2 + \tilde \sigma^2)h}{2}
           T^a \partial^a \log \rho_t(U_t)
       + \sqrt{h}\, \tilde\sigma\, \eta(t)
     \right) U_t\,,
   \label{eq:diffusion:recursive:rev:U_t}
\end{equation}
where $\rho_t(U_t)$ denotes the PDF at time $t$,
and $\eta(t) = \eta_a(t) T^a$ is an anti-Hermitian Gaussian matrix.
The second term in the exponential contains the score function
$T^a \partial^a \log p_t(U_t)$, which is the gradient of the logarithmic density.
This term drives the system back toward regions of higher probability, which causes the time-reversed process to correctly reconstruct the data distribution.



\bibliographystyle{apsrev4-1}
\bibliography{references.bib}

\end{document}